\documentstyle[12pt]{article}
\title{Modified Symplectic Structures in Cotangent Bundles of Lie Groups.}
\author{F.J. Vanhecke, C. Sigaud and A.R. da Silva \\ Instituto de F\'{\i}sica, Instituto de Matem\'{a}tica,\\ UFRJ, Rio de Janeiro, Brazil }
\date{} 
\def\be{\begin{equation}}
\def\ee{\end{equation}}
\def\eea{\end{eqnarray}}
\def\bea{\begin{eqnarray}}
\def\bean{\begin{eqnarray*}}
\def\eean{\end{eqnarray*}}
\def\bc{\begin{center}}
\def\ec{\end{center}}
\def\bA{{\bf A}}
\def\bR{{\bf R}}
\def\bX{{\bf X}}

\def\bX{{\bf X}}
\def\bY{{\bf Y}}
\def\bZ{{\bf Z}}

\def\bK{{\bf K}}
\def\bL{{\bf L}}
\def\bAd{{\bf Ad}}
\def\bad{{\bf ad}}
\def\dif{{\bf d}}

\def\bu{{\bf u}}
\def\bv{{\bf v}}
\def\bw{{\bf w}}
\def\bk{{\bf k}}

\def\bfe{{\bf e}}
\def\b1{{\bf\LARGE 1}}
\def\ep{{\epsilon}}
\def\LG{{\cal G}}
\def\cM{{\cal M}}
\def\cQ{{\cal Q}}

\def\cQ{{\cal Q}}
\def\cI{{\cal I}}

\def\cV{{\cal V}}

\begin{document}
\maketitle
\begin{abstract}
In earlier work \cite{vanhecke}, we studied an extension of the canonical symplectic structure in the cotangent bundle of an affine space $\cQ=\bR^N$, by additional terms implying the Poisson non-commutativity of both configuration and momentum variables. In this article, we claim that such an extension can be done consistently when $\cQ$ is a Lie 
group $G$. 
\end{abstract}
\section{Introduction}\label{un}
As applied to physics, noncommutative geometry is understood mainly in two ways. The first one is the spectral triple approach of A.Connes \cite{Connes} with the Dirac operator playing a central role in unifying, through the universal action principle, gravitation with the standard model of fundamental interactions. The second one is the quantum field theory on noncommutative spaces \cite{Douglas} with the Moyal product as main ingredient. Besides these, a proposition by several authors \cite{Duval,Horvathy} was made to generalise quantum mechanics in such a way that the operators corresponding to space coordinates no longer commute : $[\widehat{x}^k,\widehat{x}^\ell]\not=0\,$. 
This was implemented by an extension of the Poisson structure on the cotangent space such that the brackets satisfy $\left\{x^k,x^\ell\right\}\not=0$. 
Upon quantisation, the corresponding operators should then also be noncommutative.  
A particle moving in an affine space $\bA^N$, has its configuration, in a fixed reference frame, given by an element $\{x^k\}$ of the translation group : $\cQ=\bR^N$ with cotangent bundle $T^\star(\cQ)=\bR^N\times\bR^N$. In \cite{vanhecke}, we examined such an extension of the canonical symplectic two-form $\omega_0=dx^i\wedge dp_i\rightarrow
\Omega=\omega_0+\omega_F+\omega_B$ :
\be\label{een1}
\omega_F=\frac{1}{2}\,F_{ij}(x)\,dx^i\wedge dx^j\;,\;
\omega_B=\frac{1}{2}\,B^{k\ell}(p)\,dp_k\wedge dp_\ell\ee
This extension is form-invariant under a change of the reference frame lifted to the cotangent bundle : 
\be\label{een2}
T^\star(\cQ)\rightarrow T^\star(\cQ):
\left(x^i,p_k\right)\rightarrow\left(x^{\prime\,i}={A^i}_j\,x^j+a^k\,,\,
p^\prime_k=p_\ell\,{(A^{-1})^\ell}_k\right)\ee
\bea\label{een3}
\Omega&\rightarrow&\Omega^\prime=dx^{\prime\,i}\wedge dp_i^\prime
+\frac{1}{2}\,F_{ij}^{\,\prime}(x^\prime)\,dx^{\prime\,i}\wedge dx{\prime\,^j}
+\frac{1}{2}\,B^{\,\prime\,k\ell}(p^\prime)\,dp_k^\prime\wedge dp_\ell^\prime\nonumber\\
&&\\
&&F_{ij}^{\,\prime}(x^\prime)=F_{k\ell}(x)\,{(A^{-1})^k}_i\,
{(A^{-1})^\ell}_j\;,\;B^{\,\prime\,k\ell}(p^\prime)={A^k}_i\,{A^\ell}_j\,B^{ij}(p)\nonumber
\eea
For a general configuration space $\cQ$, a diffeomorphism $\phi:x^i\rightarrow x^{\prime\,i}\doteq\phi^i(x)$, when lifted to $T^\star(\cQ)$, becomes 
\bean
&&\widetilde{\phi}:\left(x^i,p_k\right)\rightarrow
\left(x^{\prime\,i}=\phi^i(x),p_k^\prime=p_\ell\,
\frac{\partial(\phi^{-1}(x^\prime))^\ell}{\partial x^{\prime\,k}}\right)\\
&&
F_{ij}^{\,\prime}(x^\prime)=F_{k\ell}(x)\;
\frac{\partial(\phi^{-1})^k(x^\prime)}{\partial x^{\prime\,i}}\;
\frac{\partial(\phi^{-1})^\ell(x^\prime)}{\partial x^{\prime\,j}}\\
&& B^{\,\prime\,k\ell}(p^\prime,x^\prime)=
\frac{\partial \phi^k(x)}{\partial x^i}\;\frac{\partial \phi^\ell(x)}{\partial x^j}\;
B^{ij}(p)\eean
In general $B^{\,\prime\,k\ell}$ is function of both variables $\{p^\prime,x^\prime\}$ and no 
intrinsic meaning can be given to the particular form of the extension $\Omega$ 
in equation (\ref{een1}).\\
In this work, we show that such an extension is achieved when $\cQ=G$ is a Lie group. This 
is possible because the cotangent bundle $T^\star(G)$ has two distinguished trivialisations, the left- and right trivialisations \cite{Abraham} implemented respectively by the bases of the left- and right invariant differential forms. \\
In section {\bf\ref{deux}.}, inspired by the rigid body motion, we use the left trivialisation with left invariant or {\it body-coordinates} and construct a left invariant two-form.
In the case of constant $F_{ij}$ and $B^{k\ell}$ fields the $\omega_F$ term arises from a symplectic one-cocycle, as introduced by Souriau 
\cite{Souriau,Paulette}, and $\omega_B$ will be automatically left invariant.  The constructed two-form $\Omega$ is obviously closed but the non degeneracy condition 
leads in general to a constrained Hamiltonian system.
This is examined in more detail for $SU(2)$ in section {\bf\ref{trois}.}. Final considerations are made in section {\bf\ref{quatre}.}. Some elements of Lie algebra cohomology \cite{Paulette,Azcar} are recalled in the appendix. 
\section{The phase space $\{\cM_0\equiv T^\star(G),\,\omega_0\}$}\label{deux}
\setcounter{equation}{0}
Let $\{g^\alpha\,,\,\alpha=1,2,\cdots,N\}$ be coordinates of a group element $g\in G$. Natural or holonomic coordinates of points $(g,{\bf p}_g)\in T^\star(G)$ are obtained using the basis 
$\{\dif g^\mu\}$ of the cotangent space $T_g^\star(G)$. They are given by 
$(g^\alpha,p_\mu)_{hol}$, where ${\bf p}_g=p_\mu\;\dif g^\mu$. Given a pair of dual bases $\{\bfe_\alpha\}$ of the Lie algebra $\LG\doteq T_e(G)$ and $\{\ep^\alpha\}$ of its dual $\LG^\star$, the differential and pull-back of the left- and right translations $(L_g,R_g)$ define left- and right invariant vector fields and one forms : 
$\bfe_\alpha^L(g)\doteq L_{g*|e}\,\bfe_\alpha\;$, 
$\bfe_\alpha^R(g)\doteq R_{g*|e}\,\bfe_\alpha\;$, 
$\ep^\alpha_L(g)\doteq L^*_{g^{-1}|g}\,\ep^\alpha\;$, 
$\ep^\alpha_R(g)\doteq R^*_{g^{-1}|g}\,\ep^\alpha$.
With canonical group coordinates, in terms of 
${L^\alpha}_\beta(g,h)\doteq{\partial(g\,h)^\alpha}/{\partial g^\beta}$ 
and ${R^\alpha}_\beta(g,h)\doteq{\partial(h\,g)^\alpha}/{\partial g^\beta}$, 
they are explicitely given by :
\bea\label{twee1}
\bfe_\alpha^L(g)=
{L^\mu}_\alpha(g,e)\,\frac{\partial}{\partial g^\mu}&,&
\bfe_\alpha^R(g)=
{R^\mu}_\alpha(g,e)\,\frac{\partial}{\partial g^\mu}\nonumber\\
&&\\
\ep^\alpha_L(g)=
{L^\alpha}_\mu(g^{-1},g)\,\dif g^\mu&,&
\ep^\alpha_R(g)=
{R^\alpha}_\mu(g^{-1},g)\,\dif g^\mu\nonumber
\eea
These bases implement canonical trivialisations of the tangent and cotangent bundle. 
For the cotangent bundle, which is the arena of symplectic or Hamiltonian formalism, we have 
a left and a right trivialisation : 
\bean
\lambda&:&T^\star(G)\rightarrow G\times\LG^\star:
(g,p_g=p_\mu\,\dif g^\mu)\rightarrow\left(g,\pi^L=L^*_{g|e}\,p_g=
\pi^L_\mu\,\ep^\mu\right)\\
&&\pi^L_\mu=\langle p_g,\bfe_\mu^L\rangle=p_\nu\,{L^\nu}_\mu(g,e)\\
\rho&:&
T^\star(G)\rightarrow G\times\LG^\star:
(g,p_g=p_\mu\,\dif g^\mu)\rightarrow\left(g,\pi^R=R^*_{g|e}\,p_g=
\pi^R_\mu\,\ep^\mu\right)\\
&&\pi^R_\mu=\langle p_g,\bfe_\mu^R\rangle=p_\nu\,{R^\nu}_\mu(g,e)
\eean
They can be viewed as a change of coordinates of a point $(g,p_g)$ in 
$T^\star(G)$ :
\be\label{twee2}
(g,{\bf p}_g)\leftrightarrow(g^\alpha,p_\mu)_{hol}\leftrightarrow(g^\alpha,\pi^L_\mu)_{\bf B}\leftrightarrow(g^\alpha,\pi^R_\mu)_{\bf S}\ee
In rigid body theory, the coordinates of the left trivialisation are the "body" coordinates, whence the subscript $(\,,)_{\bf B}$. The right trivialisation yields "space" coordinates with subscript $(\,,)_{\bf S}$. Both are related through the coadjoint representation of $G$ in $\LG^\star$ : 
\be\label{twee3}
\pi^R_\mu={{\bf K}_\mu}^\nu(g)\;\pi_\nu^L={{\bf Ad}^\nu}_\mu(g^{-1})\,\pi_\nu^L\ee
Lifting the left multiplication in $G$ to the cotangent bundle yields a group action :
$\widetilde{L}_a:T^\star(G)\rightarrow T^\star(G):
x=(g,p_g)\rightarrow y=(ag,p^{\,\prime}_{ag}=L^\star_{a^{-1}|ag}\,p_g)$. 
In body coordinates : 
$\left(\widetilde{L}_a\right)_{\bf B}:(g^\alpha,\pi_\mu^L)_{\bf B}\rightarrow 
((ag)^\alpha,\pi_\mu^L)_{\bf B}$. The pull-back of the cotangent projection $\kappa:T^\star(G)\rightarrow G:x\doteq(g,p_g)\rightarrow g$, acting on the $\{\ep^\alpha(g)\}$ yield $\widetilde{L}_a$ invariant 
one forms on $T^\star(G)$ : 
$\langle\ep^\alpha_L(x)|=\kappa_x^\star\;\ep^\alpha_L(\kappa(x))$ and the differentials of the left invariant functions $\pi_\mu^L$ on $T^\star(G)$ also yield 
$\widetilde{L}_a$ invariant one forms on $T^\star(G)$. 
Together they provide a left invariant basis of the cotangent space at 
$x=(g^\alpha,\pi^L_\mu)_{\bf B}\in T^\star(G)$ : 
\be\label{twee4}\left\{\langle\ep^\alpha_L|\doteq{L^\alpha}_\mu(g^{-1},g)\,\langle\dif g^\mu|
\;,\;\langle\ep^L_\mu|\doteq\langle \dif\pi^L_\mu|\right\}\ee 
Its dual basis in the tangent space $T_x(T^\star(G))$ is given by 
\be\label{twee5}\left\{
|\bfe^L_\alpha\rangle\doteq|{\partial}/{\partial g^\mu}\rangle\;{L^\mu}_\alpha(g,e)
\;,\;|\bfe_L^\mu\rangle\doteq|\partial/\partial \pi^L_\mu\rangle\right\}\ee
The canonical Liouville one-form $\langle\theta_0|=p_\alpha\;\langle dg^\alpha|$ and its associated symplectic two-form 
$\omega_0=-\dif \theta_0=\langle\dif g^\alpha|\wedge \langle\dif p_\alpha|$, 
are obtained as :
\be\label{twee6}
\langle\theta_0|=\pi_\mu^L\;\langle\ep^\mu_L|\;,\;
\omega_0=\langle\ep^\mu_L|\wedge\langle\ep_\mu^L| +
\frac{1}{2}\,\pi_\mu^L\;{{\bf f}^\mu}_{\alpha\beta}\;
\langle\ep^\alpha_L|\wedge\langle\ep^\beta_L|
\ee
The Hamiltonian vector field associated to a function $A(g,\pi^L)$ on phase space 
$\cM_0\equiv T^\star(G)$, is defined by : $\imath_{\,\bX}\,\omega_0=\langle\dif A|\,$. 
Its components are :
\bea\label{twee7}
X^\mu&\doteq&\langle\ep^\mu_L|\bX\rangle=\langle \dif A|\bfe^\mu_L\rangle\nonumber\\
X_\alpha&\doteq&\langle\ep^L_\alpha|\bX\rangle=-\langle\dif A|\bfe^L_\alpha\rangle-\pi^L_\mu\,{{\bf f}^\mu}_{\alpha\beta}\,
\langle \dif A|\bfe^\beta_L\rangle
\eea
With $\imath_{\,\bY}\,\omega_0=\langle\dif B|\,$, the Poisson bracket of dynamical variables : 
$\{A,B\}_0\doteq\,\omega_0\left(\bX,\bY\right)$, is obtained explicitely in 
$(g^\alpha,\pi^L_\mu)$ variables as : 
\be\label{twee8}
\{A,B\}_0=
\langle\dif A|\bfe^L_\alpha\rangle
\;\frac{\partial B}{\partial \pi^L_\alpha}
-\frac{\partial A}{\partial \pi^L_\alpha}\;
\langle\dif B|\bfe^L_\alpha\rangle
-\frac{\partial A}{\partial \pi^L_\alpha}\;\pi^L_\mu\;
{{\bf f}^\mu}_{\alpha\beta}\;\frac{\partial B}{\partial \pi^L_\beta}
\ee
In particular, the basic Poisson brackets are :
\bea\label{twee9}
\left\{g^\alpha,g^\beta\right\}_0=0&,&
\left\{g^\alpha,\pi_\nu^L\right\}_0={L^\alpha}_\nu(g,e)\nonumber\\
\left\{\pi_\mu^L,g^\beta\right\}_0=-{L^\beta}_\mu(g,e)&,&
\left\{\pi_\mu^L,\pi^L_\nu\right\}_0=-\,\pi^L_\kappa\;{{\bf f}^\kappa}_{\mu\nu}
\eea
The flow of a particular observable, the Hamiltonian $H(g,\pi^L)$, determines the time evolution of any observable $A(g,\pi^L)$ by the equation : 
$dA/dt=\{A,H\}_0$. We assume a Hamiltonian is of the form $H(g,\pi^L)=K(\pi^L)+V(g)$.\\
Here, as in rigid body mechanics, the {\it kinetic energy} is given by 
\be\label{twee10}
K\doteq\frac{1}{2}\,\cI^{\alpha\beta}\,\pi^L_\alpha\;\pi^L_\beta
\ee
where $\cI^{\alpha\beta}$ is the inverse of a constant, positive definite, 
{\it inertia tensor} $I_{\mu\nu}$ in the "body" frame. The {\it potential energy} is a function $V$ defined on the group manifold. 
The Euler equations of motion read : 
\bea\label{twee11}
\langle\ep^\alpha_L|dg/dt\rangle&=&
{L^\alpha}_\beta(g^{-1},g)\,\frac{d\,g^\beta}{dt}=\frac{\partial K}{\partial \pi^L_\alpha}\\
\label{twee12}
\langle\ep^L_\mu|d\pi^L/dt\rangle&=&
\frac{d\,\pi^L_\mu}{dt}=-\,\frac{\partial V}{\partial g^\alpha}\,{L^\alpha}_\mu(g,e)
+\frac{\partial K}{\partial \pi^L_\nu}\;\pi^L_\alpha\,{{\bf f}^\alpha}_{\nu\mu}
\eea
The first of these equations {\bf(\ref{twee11})} relates the angular momentum $\pi^L_\alpha$ with the angular velocity in the body frame $\Omega^\mu_L$  : 
\be\label{twee13}
\Omega^\alpha_L\doteq{L^\alpha}_\beta(g^{-1},g)\,\frac{dg^\beta}{dt}=
\cI^{\alpha\mu}\,\pi^L_\mu\;;\;
\pi_\mu^L=I_{\mu\nu}\,\Omega^\nu_L\ee
while the second {\bf(\ref{twee12})} takes the classical form 
\be\label{twee14}
\frac{d\pi^L_\mu}{dt}+\pi^L_\kappa\;{{\bf f}^\kappa}_{\mu\nu}\,\Omega^\nu_L=-\,\frac{\partial V}{\partial g^\alpha}\,{L^\alpha}_\mu(g,e)\ee
An example of $V(g)$ is given by a {\it gravitational potential energy} as follows.
Let $\bL=\bfe_\alpha\,L^\alpha$ be a constant vector in $\LG$ (the position of the centre of mass in the body frame) and $\gamma=\gamma_\alpha\,\ep^\alpha$ a constant vector in $\LG^\star$ (the gravitational force in the space fixed frame). 
The potential energy is defined as :
\be\label{twee15}
V(g)\doteq -\,\left(\gamma\,|\,\bAd(g)\,\bL\right)=-\,\left(\bK(g^{-1})\gamma\,|\,\bL\right)\ee
where $(\;|\;)$ denotes the canonical pairing between $\LG$ and its dual $\LG^\star$. 
To compute $\langle\dif V|\bfe_\mu^L\,\rangle$ we use the representation of the Maurer-Cartan form :
\[D(g^{-1})\,\dif D(g)=D^\prime(g^{-1}\,\dif g)\]
where $D$ is any representation $D$ of $G$, with derived representation $D^\prime$ of $\LG$. In particular,  
$\dif \bAd(g)=\bAd(g)\,\bad(\bfe_\mu)\,\ep_L^\mu(g)$ and 
$\dif \bK(g)=\bK(g)\,\bk(\bfe_\mu)\,\ep_L^\mu(g)$. This yields :
\be\label{twee16}
\langle\dif V|\bfe_\mu^L\,\rangle(g)=
-\,\left(\bK(g^{-1})\,\gamma\,|\,\bad(\bfe_\mu)\,\bL\right)
=-\,\left(\Gamma(g)\,|\,\bad(\bfe_\mu)\,\bL\right)\ee
where $\Gamma(g)\doteq\bK(g^{-1})\,\gamma$ is the variable gravitational force in the body-fixed frame. Using the above formulae to compute $\dif\bK(g^{-1})$, we obtain :
\be\label{twee17}
\frac{d\,\Gamma_\mu}{dt}=\left(\Gamma\,|\,\bad(\bfe_\mu)\,\Omega_L\right)=\Gamma_\alpha\,{{\bf f}^\alpha}_{\mu\beta}\,\Omega_L^\beta\ee
Equation {\bf(\ref{twee14})} reads : 
\be\label{twee18}
\frac{d\pi^L_\mu}{dt}+\pi^L_\alpha\;{{\bf f}^\alpha}_{\mu\beta}\,\Omega^\beta_L=
\left(\Gamma\,|\,\bad(\bfe_\mu)\,\bL\right)=\Gamma_\alpha\,{{\bf f}^\alpha}_{\mu\beta}\,
L^\beta\ee
Together with {\bf(\ref{twee13})},
\[\Omega^\alpha_L\doteq{L^\alpha}_\beta(g^{-1},g)\,\frac{dg^\beta}{dt}=
\cI^{\alpha\mu}\,\pi^L_\mu\]
the equations {\bf(\ref{twee17})} and {\bf(\ref{twee18})}
form the so-called Euler-Poisson system.
\section{Modified symplectic structure on $T^\star(G)$}\label{trois}
\setcounter{equation}{0}
In appendix {\bf\ref{sept}} it is shown that, if $\Theta=\frac{1}{2}\,\Theta_{\alpha\beta}\,\ep^\alpha\wedge\ep^\beta\in\Lambda^2(\LG^\star)$, obeys the cocycle condition 
{\bf(\ref{zeven1})}, then  
$\Theta_L(g)\doteq(1/2)\,\Theta_{\alpha\beta}\;\ep^\alpha_L(g)\wedge\ep^\beta_L(g)$
is a closed left-invariant two-form on $G$. Including this closed two-form in the canonical two-form, one obtains another symplectic two-form on $T^\star(G)$, which, furthermore, is $\widetilde{L}_a$ invariant. So we define :
\be\label{drie1}
\omega_I=\omega_0-\Theta_L=
\langle\ep_L^\mu|\wedge\langle\dif\pi^L_\mu|+
\frac{1}{2}\,\left(\pi_\mu^L\;{{\bf f}^\mu}_{\alpha\beta}-\Theta_{\alpha\beta}\right)\;
\langle\ep_L^\alpha|\wedge\langle\ep_L^\beta|
\ee
The Poisson brackets are also modified and {\bf(\ref{twee8})}, {\bf(\ref{twee9})} become :
\bea\label{drie2}
\{A,B\}_I&=&
\frac{\partial A}{\partial g^\mu}\,{L^\mu}_\alpha(g,e)
\,\frac{\partial B}{\partial \pi^L_\alpha}
\,-\,
\frac{\partial B}{\partial g^\mu}\,{L^\mu}_\alpha(g,e)
\,\frac{\partial A}{\partial \pi^L_\alpha}\nonumber\\
&&
-\;\left(\pi^L_\mu\;{{\bf f}^\mu}_{\alpha\beta}-\Theta_{\alpha\beta}\right)\;
\frac{\partial A}{\partial \pi^L_\alpha}\;\frac{\partial B}{\partial \pi^L_\beta}
\eea
In particular, the fundamental brackets are :
\bea\label{drie3}
\left\{g^\alpha,g^\beta\right\}_I=0&,&
\left\{g^\alpha,\pi_\nu^L\right\}_I={L^\alpha}_\nu(g,e)\nonumber\\
\left\{\pi_\mu^L,g^\beta\right\}_I=-{L^\beta}_\mu(g,e)&,&
\left\{\pi_\mu^L,\pi^L_\nu\right\}_I=-\,\left(\pi^L_\kappa\;{{\bf f}^\kappa}_{\mu\nu}
-\Theta_{\mu\nu}\right)
\eea
The modified symplectic structure induces an additional interaction and the Euler equations 
become : 
\bea\label{drie4}
\Omega^\alpha_L\doteq{L^\alpha}_\beta(g^{-1},g)\,
\frac{d g^\beta}{dt}&=&\frac{\partial K}{\partial \pi^L_\alpha}=\cI^{\alpha\mu}\,\pi^L_\mu
\\
\label{drie5}
\frac{d\pi^L_\mu}{dt}&=&-\langle\dif V|\bfe_\mu^L\rangle+
\frac{\partial K}{\partial\pi^L_\alpha}\,\left(\pi^L_\kappa\;{{\bf f}^\kappa}_{\alpha\mu}
-\Theta_{\alpha\mu}\right)\eea
The relation between the velocity in the body frame and the angular 
momentum {\bf(\ref{twee13})} is maintained :
$\pi_\mu^L=I_{\mu\nu}\;\Omega^\nu_L$, 
while the second {\bf(\ref{twee14})} takes the  interaction into account : 
\be\label{drie6}
\frac{d\pi^L_\mu}{dt}+\pi^L_\kappa\;{{\bf f}^\kappa}_{\mu\alpha}\,\Omega_L^\alpha\,=
-\langle\dif V|\bfe_\mu^L\rangle-
\Omega_L^\alpha\;\Theta_{\alpha\mu}\ee
For a semisimple Lie algebra $\LG$, we have $\Theta_{\alpha\beta}=-\,\xi_\mu\;{{\bf f}^\mu}_{\alpha\beta}$ and we may define a modified Liouville one-form : 
\be\label{drie7}
\langle\theta_I|=\pi^\prime_\mu\;\langle\ep^\mu_L|\;,\;
\pi_\mu^\prime\doteq\pi^L_\mu+\xi_\mu\ee
and the symplectic two-form reads
\be\label{drie8}
\omega_I=\,-\,\dif\langle\theta_I\,|=\langle\ep_L^\mu|\wedge\langle\dif\pi^\prime_\mu|+
\frac{1}{2}\,\pi^\prime_\mu\;{{\bf f}^\mu}_{\alpha\beta}\;
\langle\ep_L^\alpha|\wedge\langle\ep_L^\beta|
\ee
This means that 
such that $\{g^\alpha,{p^\prime}_\mu=p_\mu+\xi_\beta\,{L^\beta}_\mu(g^{-1};g)\}$ are Darboux coordinates :
\be\label{drie9}
\langle\theta_I|={p^\prime}_\mu\;\langle\dif g^\mu|\;,\;\omega_I\doteq-\dif\langle\theta_I|=\langle\dif g^\mu|\wedge
\langle\dif{p^\prime}_\mu|\ee
In $\left(g^\alpha,\pi^\prime_\mu\right)$ coordinates, the Hamiltonian reads
\be\label{drie10}
H^\prime=K^\prime(\pi^\prime)+V(g)=\frac{1}{2}\,\cI^{\mu\nu}\,(\pi^\prime_\mu-\xi_\mu)\,(\pi^\prime_\nu-\xi_\nu)+V(g)
\ee
and the Euler equations read :
\bea\label{drie11}
{L^\alpha}_\beta(g^{-1},g)\,
\frac{d g^\beta}{dt}&=&\frac{\partial K^\prime}{\partial \pi^\prime_\alpha}=\cI^{\alpha\mu}\,(\pi^\prime_\mu-\xi_\mu)
\\
\label{drie5}
\frac{d\pi^\prime_\mu}{dt}&=&-\langle\dif V|\bfe_\mu^L\rangle+
\frac{\partial K^\prime}{\partial\pi^\prime_\alpha}\,
\left(\pi^\prime_\kappa\;{{\bf f}^\kappa}_{\alpha\mu}\right)\eea
which, obviously are equivalent to {\bf(\ref{drie4})} and {\bf(\ref{drie5})}.
\section{The closed two-form $\omega_L$}\label{quatre}
\setcounter{equation}{0}
Configuration space coordinates which do not Poisson commute, are obtained through the addition of a left-invariant and closed two-form to {\bf(\ref{drie1})} :
\be\label{vier1}
\Upsilon^L\doteq\frac{1}{2}\,\Upsilon^{\mu\nu}\,
\langle\dif\pi^L_\mu|\wedge\langle\dif\pi^L_\nu|\ee
\bea\label{vier2}
\omega_L\doteq\omega_0-\Theta_L+\Upsilon^L&=&
\langle\ep_L^\mu|\wedge\langle\dif\pi^L_\mu|
+
\frac{1}{2}\,\left(\pi_\mu^L\;{{\bf f}^\mu}_{\alpha\beta}-\Theta_{\alpha\beta}\right)\;
\langle\ep_L^\alpha|\wedge\langle\ep_L^\beta|\nonumber\\
&&+
\frac{1}{2}\,\Upsilon^{\mu\nu}\,
\langle\dif\pi^L_\mu|\wedge\langle\dif\pi^L_\nu|
\eea
With the notation $S_{\alpha\beta}\equiv\left(\pi_\mu^L\;{{\bf f}^\mu}_{\alpha\beta}-
\Theta_{\alpha\beta}\right)$, we wite $\omega_L$ in matrix form :
\be\label{vier3}
\omega_L\equiv
\frac{1}{2}
\left( \langle\ep_L^\alpha|\quad\langle\dif\pi^L_\mu|\right)\,\wedge\,
\left(\begin{array}{cc}
S_{\alpha\beta} & {\delta_\alpha}^\nu\\
&\\
-{\delta^\mu}_\beta & \Upsilon^{\mu\nu}\end{array}\right)\,
\left(\begin{array}{c} \langle\ep_L^\beta| \\ \\ \langle\dif\pi_\nu^L|\end{array}\right)\ee
The degeneracy of $(\omega_L)$ is examined comsidering the equation 
\be\label{vier4}
\imath_{|\bX\rangle}\omega_L=\langle\dif A|\ee
In the bases {{\bf(\ref{twee4})}, {\bf(\ref{twee5})}: 
$X^\alpha\doteq\langle\ep^\alpha_L|\bX\rangle\;,\;X_\mu\doteq\langle\ep_\mu^L|\bX\rangle$
and {\bf(\ref{vier4})} reads :
\be\label{vier5}
X^\alpha\,{\Phi_\alpha}^\nu
=\langle\dif A|\bfe^\nu_L\rangle+\langle\dif A|\bfe^L_\mu\rangle\Upsilon^{\mu\nu}\,,\,
X_\mu\,{\Psi^\mu}_\beta
=-\langle\dif A|\bfe^L_\beta\rangle+\langle\dif A|\bfe^\alpha_L\rangle S_{\alpha\beta}\ee
where we introduced the matrices, linear in the momenta :  
\be\label{vier6}
{\Phi_\alpha}^\nu\doteq {\delta_\alpha}^\nu+S_{\alpha\mu}\Upsilon^{\mu\nu}\;,\;
{\Psi^\mu}_\beta\doteq{\delta^\mu}_\beta+\Upsilon^{\mu\nu}S_{\nu\beta}\ee
They are mutually transposed and the products  
$\Phi\,S=S\,\Psi\,,\,\Upsilon\,\Phi=\Psi\,\Upsilon$ are antisymmetric. 
The fundamental equation {\bf(\ref{vier4})}, defining Hamiltonian vector fields, 
has a solution if $\Phi$ and $\Psi$ have inverses, i.e. if
\be\label{vier7}\Delta\doteq
\det\Phi\equiv\det\Psi\not=0\ee
The matrices $\Upsilon\,\Phi^{-1}=\Psi^{-1}\,\Upsilon$ and 
$\Phi^{-1}\,S=S\,\Psi^{-1}$ are then also antisymmetric. The Hamiltonian vector fields are obtained as :
\bea\label{vier8}
X^\alpha&=&{(\Psi^{-1})^\alpha}_\mu\,\left(\langle\dif A|\bfe^\mu_L\rangle
-\Upsilon^{\mu\nu}\,\langle\dif A|\bfe_\nu^L\rangle\right)\,
\nonumber\\
&=&\left(\langle\dif A|\bfe^\nu_L\rangle+\langle\dif A|\bfe_\mu^L\rangle\,\Upsilon^{\mu\nu}\right)\,{(\Phi^{-1})_\nu}^\alpha
\nonumber\\
X_\mu&=&{(\Phi^{-1})_\mu}^\alpha\,\left(-\langle\dif A|\bfe_\alpha^L\rangle-
S_{\alpha\beta}\,\langle\dif A|\bfe^\beta_L\rangle\,\right)\nonumber\\
&=&\left(-\langle\dif A|\bfe_\beta^L\rangle+
\langle\dif A|\bfe^\alpha_L\rangle\,S_{\alpha\beta}\right)\,{(\Psi^{-1})^\beta}_\mu\eea
The Poisson brackets between the basic dynamical variables are :
\bea\label{vier9}
&&\left\{g^\alpha,g^\beta\right\}_L=
-\,{L^\alpha}_\kappa(g,e)\,{L^\beta}_\lambda(g,e)\;\Upsilon^{\kappa\mu}\,
{(\Phi^{-1})_\mu}^\lambda\nonumber\\
&&
\left\{g^\alpha,\pi_\nu^L\right\}_L=
{L^\alpha}_\kappa(g,e)\,{(\Psi^{-1})^\kappa}_\nu\;,\;
\left\{\pi_\mu^L,g^\beta\right\}_L=
-\,{L^\beta}_\kappa(g,e)\,{(\Psi^{-1})^\kappa}_\mu\nonumber\\
&&\left\{\pi_\mu^L,\pi_\nu^L\right\}_L=
-\,S_{\mu\kappa}\,{(\Psi^{-1})^\kappa}_\nu
\eea
For a Hamiltonian $H=K+V$, the equations of motion are :
\bea\label{vier10}
\Omega^\alpha_L\doteq{L^\alpha}_\beta(g^{-1},g)\,
\frac{d g^\beta}{dt}&=&\left(\frac{\partial K}{\partial \pi^L_\nu}
+\langle\dif V|\bfe_\mu^L\rangle\,\Upsilon^{\mu\nu}\right)\,{(\Phi^{-1})_\nu}^\alpha
\\
\label{vier11}
\frac{d\pi^L_\mu}{dt}&=&\left(-\langle\dif V|\bfe_\beta^L\rangle+
\frac{\partial K}{\partial\pi^L_\alpha}\,S_{\alpha\beta}\right)\,{(\Psi^{-1})^\beta}_\mu
\eea
Since $\Phi\,,\,\Psi$ are linear in $\pi^L$, $\Delta$ is a polynomial in $\pi^L$ of degree at most equal to $N$, the dimension of the Lie group.
It defines an algebraic variety in $\LG^\star$ : 
\be\label{vier12}\Pi_1\doteq\{(g,\pi^L)|\Delta(\pi^L)=0\}\ee
and its complement $\cV_\Delta\doteq\LG^\star\backslash\Pi_1$ defines a manifold 
\be\label{vier13}
\cM_0^\prime\doteq G\times\cV_\Delta\ee
with symplectic structure given by $\omega_L$, restricted to $\cM_0^\prime$. If it happens that $\Pi_1$ itself is an algebraic manifold, an imbedded submanifold is obtained :
\be\label{vier14}\cM_1\doteq G\times\Pi_1\ee
with imbedding in $\cM_0\doteq G\times\LG^\star$ :
$j_1:\cM_1\hookrightarrow \cM_0$. The system is then constrained to $\cM_1$ and we may 
look for solutions of {\bf(\ref{vier4})} restricted to $\cM_1$. 
Such solutions may exist if further conditions are imposed on the Hamiltonian. To proceed systematically, we follow the algorithm of Gotay, Nester and Hinds \cite{GNH}. To keep things simple, this will be done in the next section for the semi-simple group $SU(2)$.
\section{A case study : $SU(2)$}\label{cinq}
\setcounter{equation}{0}
The dynamical variables are functions on $\cM_0\doteq SU(2)\times su(2)^\star$. A basis 
$\{\bfe_\alpha\}$ of the Lie algebra $su(2)$ may be chosen such that its structure constants  are the Kronecker symbols $[\bfe_\alpha,\bfe_\beta]=\bfe_\mu\,{\ep^\mu}_{\alpha\beta}$. The Killing metric 
$\eta_{\alpha\beta}\doteq {\ep^\mu}_{\alpha\nu}\,{\ep^\nu}_{\beta\mu}=
-2\,\delta_{\alpha\beta}$, provides an isomorphism between $su(2)$ and $su(2)^\star$. The metric $\delta_{\alpha\beta}$ with inverse 
$\delta^{\mu\nu}$ will be freely used to raise or to lower indices. 
$\Theta_L$ is written in terms of a {\it magnetic field} $\xi_\mu$ as $\Theta_{\alpha\beta}=-\xi_\kappa\, {\ep^\kappa}_{\alpha\beta}$ and any antisymmetric $\Upsilon$ can be written in terms of $\tau^\lambda$, a {\it dual magnetic field in momentum space}, as 
$Y^{\mu\nu}=\tau^\lambda\,{\ep_\lambda}^{\mu\nu}$. Defining $\pi^\prime_\kappa\doteq\pi^L_\kappa+\xi_\kappa$, $\omega_L$ reads :
\be\label{vijf1}
\omega_L\equiv
\frac{1}{2}
\left( \langle\ep_L^\alpha|\quad\langle\dif\pi^L_\mu|\right)\,\wedge\,
\left(\begin{array}{cc}
\pi^\prime_\kappa\, {\ep^\kappa}_{\alpha\beta} & {\delta_\alpha}^\nu\\
&\\
-{\delta^\mu}_\beta & \tau^\lambda\,{\ep_\lambda}^{\mu\nu}\end{array}\right)\,
\left(\begin{array}{c} \langle\ep_L^\beta| \\ \\ \langle\dif\pi_\nu^L|\end{array}\right)\ee
The fundamental equation {\bf(\ref{vier4})} :
$\imath_{\,|{\bf X}\rangle}\,\omega_L=\langle\dif H|$ becomes : 
\[
X^\alpha\,\pi^\prime_\kappa\,{\ep^\kappa}_{\alpha\beta}-X_\beta=H_\beta\;,\;
X^\nu+X_\mu\,\tau^\lambda\,{\ep_\lambda}^{\mu\nu}=H^\nu\]
where $H_\beta\doteq\left(\partial H/\partial g^\alpha\right)\,{L^\alpha}_\beta(g,e)\;,\;
H^\nu\doteq\left(\partial H/\partial\pi^L_\nu\right)$. The matrices {\bf(\ref{vier6})} are 
given explicitely by 
${\Phi_\alpha}^\nu\doteq
C_1\,{\delta_\alpha}^\nu+\tau_\alpha\pi^{\prime\,\nu}$ and 
${\Psi^\mu}_\beta\doteq C_1\,{\delta^\mu}_\beta+\pi^{\prime\,\mu}\tau_\beta$, where 
$C_1\doteq (1-\pi^\prime\cdot\tau)$. They obey ${\Phi_\alpha}^\nu\,\left({\delta_\nu}^\beta-\tau_\nu\,\pi^{\prime\,\beta}\right)=
C_1\,{\delta_\alpha}^\beta$ and 
${\Psi^\mu}_\beta\,\left({\delta^\beta}_\nu-\pi^{\prime\,\beta}\,\tau_\nu\right)=C_1\,
{\delta^\mu}_\nu$. It follows that {\bf(\ref{vier5})} implies :
\bea\label{vijf2}
X^\alpha\,
(1-\pi^\prime\cdot\tau)&=&
H^\alpha-\pi^{\prime\,\alpha}\,(\tau_\beta\,H^\beta)-{\ep^{\alpha\mu}}_\nu\;H_\mu\;\tau^\nu
\\
\label{vijf3}
X_\mu\,(1-\pi^\prime\cdot\tau)&=&
-H_\mu+\tau_\mu\,(\pi^{\prime\,\nu}\,H_\nu)-{\ep_{\mu\alpha}}^\beta\;H^\alpha\;\pi^\prime_\beta
\eea
\subsection{The non degenerate case}
The determinant of the matrices $\Phi$ and $\Psi$ is given by 
$\Delta=(C_1)^2$. 
Obviously the plane $\Pi_1\doteq\{(g,\pi^L)|(1-\pi^\prime\cdot\tau)=0\}$ is an algebraic manifold in $\LG^\star$.  Its complement $\cV_\Delta\doteq\LG^\star\backslash\Pi_1$ 
defines a manifold $\cM_0^\prime\doteq G\times\cV_\Delta$ with symplectic structure 
$\omega_L$, retricted to $\cM_0^\prime$. 
On $\cM_0^\prime$, $\Phi$ and $\Psi$ have inverses :
\[{(\Psi^{-1})^\beta}_\nu=(C_1)^{-1}\,
\left({\delta^\beta}_\nu-\pi^{\prime\beta}\tau_\nu\right)\;,\;
{(\Phi^{-1})_\nu}^\beta=(C_1)^{-1}\,
\left({\delta_\nu}^\beta-\tau_\nu\pi^{\prime\beta}\right)\]
For a Hamiltonian $H=K(\pi^L)+V(g)$, the Hamiltonian vector fields are read off from {\bf(\ref{vijf2})} and {\bf(\ref{vijf3})} with ensuing equations of motion :
\bea\label{vijf4}
\Omega^\alpha_L&\doteq&
{L^\alpha}_\beta(g^{-1},g)\,\frac{dg^\beta}{dt}=
\left(\frac{\partial K}{\partial\pi^L_\nu}+\langle\dif V|\bfe_\mu^L\rangle\,\tau^\lambda\,{\ep_\lambda}^{\mu\nu}\right)\,{(\Phi^{-1})_\nu}^\alpha
\nonumber\\
\frac{d\pi^L_\mu}{dt}&=&
\left(-\langle\dif V|\bfe_\beta^L\rangle+\frac{\partial K}{\partial \pi^L_\alpha}\;\pi^\prime_\kappa\,{\ep^\kappa}_{\alpha\beta}\right)\,{(\Psi^{-1})^\beta}_\mu
\eea
For a purely kinetic Hamiltonian, we obtain :
\be\label{vijf5}
\Omega^\alpha_L= \frac{\partial K}{\partial\pi^L_\mu}\,{(\Phi^{-1})_\mu}^\alpha\;,\;
\frac{d\pi^L_\mu}{dt}=\Omega_L^\alpha\,\pi^\prime_\beta\,{\ep^\beta}_{\alpha\mu}
\ee
\subsection{The degenerate case}
The equation $C_1\equiv(1-\pi^\prime\cdot\tau)=0$ defines a two dimensional plane 
$\Pi_1$ in $su(2)^\star\cong\bR^3$. The {\it primary constrained manifold}, defined by $\cM_1\doteq SU(2)\times\Pi_1$, is imbedded in $\cM_0\doteq SU(2)\times su(2)^\star$. 
On $\cM_1$, the closed two-form $\omega_L$ is degenerate and the pairing of $\pi^\prime\in su(2)^\star$ with $\tau\in su(2)$ equals 1. So $|\tau\rangle\not=0$ and, without loss of generality, we take $\{\tau^\alpha\}=\{0,0,\tau\}$. In what follows, greek indices $\{\alpha,\beta,\mu,\nu,\cdots\}$ shall vary in $\{1,2,3\}$, while latin indices $\{a,b,m,n,\cdots\}$ assume only the values $\{1,2\}$.
The imbedding is given by :
\be\label{vijf6}
j_1 : \cM_1\hookrightarrow \cM_0: x_1\equiv(g^\alpha,\pi^L_m)\rightarrow x_0=j_1(x_1)\equiv
(g^\alpha,\pi^L_m,\pi^L_3=1/\tau-\xi_3)\ee
with its differential or push-forward :
\be\label{vijf7}
j_{1\star}:T\cM_1\rightarrow T\cM_0:(x_1;X^\alpha,X_m)\rightarrow(x_0;X^\alpha,X_m,X_3=0)\ee
The pull-back transforms forms on $\cM_0$ into forms on $\cM_1$ :
\be\label{vijf8}
{j_1}^\star:{\bigwedge}^\bullet(T^\star\cM_0)\rightarrow{\bigwedge}^\bullet(T^\star\cM_1)\ee
In particular the pull-back of $\omega_L$ to the five dimensional manifold $\cM_1$ is 
\be\label{vijf9}
\widetilde{\omega}_{L|\,1}\doteq {j_1}^\star(\omega_L)\ee 
The restriction of $\omega_L$ to $\cM_1$, not to be confused with its pull-back, is denoted by $\omega_{L|\,1}\doteq\omega_L\circ j_1$. In matrix representation :
\bea\label{vijf10}
\omega_{L|\,1}&=&\frac{1}{2}\,
\left(\,\langle\ep_L^\alpha|\quad
\langle\dif\pi^L_{\mu}|\,\right)\,\wedge\,
\left(\begin{array}{cccccc}
0 & 1/\tau & -\,\pi^\prime_2     & 1 & 0 & 0\\
-1/\tau & 0 & \pi^\prime_1       & 0 & 1 & 0\\
\pi^\prime_2 & -\pi^\prime_1 & 0 & 0 & 0 & 1\\
-1 & 0 & 0 & 0     & \tau & 0\\
0 & -1 & 0 & -\tau & 0    & 0\\
0 & 0  & -1& 0     & 0    & 0
\end{array}\right)\;
\left(\begin{array}{c}
\phantom{1}\\
\langle\ep_L^\beta|\\
\phantom{3}\\
\phantom{4}\\
\langle\dif\pi^L_{\nu}| \\
\phantom{6}
\end{array}\right)\nonumber\\
&&
\eea
Let $(T\cM_0)_{|\,1}\doteq\{(x,\bX)\in T\cM_0\,|\,x\in\cM_1\}$ be the subbundle of $T\cM_0$
restricted to $\cM_1$. Following the GNH algorithm \cite{GNH}, we look for a vector field 
$|\bX\rangle$ in $(T\cM_0)_{|\,1}$, tangent to $\cM_1$ and solution of 
\be\label{vijf11}
\imath_{|\bX\rangle}\omega_{L|\,1}=\langle\dif H|\circ j_1\ee
Explicitely :
\bean
-(1/\tau)\,X^2+\pi^\prime_2\,X^3-X_1&=&\langle\dif V|\bfe^L_1\rangle\nonumber\\
+(1/\tau)\,X^1-\pi^\prime_1\,X^3-X_2&=&\langle\dif V|\bfe^L_2\rangle\nonumber\\
-\pi^\prime_2\,X^1+\pi^\prime_1\,X^2-X_3&=&\langle\dif V|\bfe^L_3\rangle\nonumber\\
X^1-\tau\,X_2&=&\partial K/\partial\pi^L_1\nonumber\\
X^2+\tau\,X_1&=&\partial K/\partial\pi^L_2\nonumber\\
X^3&=&\partial K/\partial\pi^L_3\nonumber\eean
Two independent null vectors of $\omega_{L|\,1}$, solution of 
$\imath_{|\bZ\rangle}\omega_{L|\,1}=0$, are given by :
\bea\label{vijf12}
|\bZ^1\rangle&=&|\bfe_1^L\rangle+(1/\tau)\,|\partial/\partial\pi^L_2\rangle-\pi^\prime_2\,
|\partial/\partial\pi^L_3\rangle\nonumber\\
|\bZ^2\rangle&=&|\bfe_2^L\rangle-(1/\tau)\,|\partial/\partial\pi^L_1\rangle+\pi^\prime_1\,
|\partial/\partial\pi^L_3\rangle
\eea
Consistency requires $\{\langle\dif H|\bZ^a\rangle=0\}$ for $(a=1,2)$ and 
$\pi^\prime_3=1/\tau$.
\bea\label{vijf13}
C_{21}\equiv\pi^\prime_2\,(\partial K/\partial\pi_3^L)-
\pi^\prime_3\,(\partial K/\partial\pi_2^L)-\langle\dif V|\bfe_1^L\rangle
&=&0\nonumber\\
C_{22}\equiv
\pi^\prime_3\,(\partial K/\partial\pi_1^L)-
\pi^\prime_1\,(\partial K/\partial\pi_3^L)-
\langle\dif V|\bfe_2^L\rangle&=&0
\eea
These two equations define a secondary constrained manifold $\cM_2\subset\cM_1$, 
on which a  particular solution of {\bf(\ref{vijf11})} is 
\be\label{vijf14}|\bX_P\rangle=
|\bfe_1^L\rangle\,\partial K/\partial\pi^L_1+
|\bfe_2^L\rangle\,\partial K/\partial\pi^L_2+
|\bfe_3^L\rangle\,\partial K/\partial\pi^L_3+|\partial/\partial\pi^L_3\rangle\,C_{23}\ee
where $C_{23}\equiv\pi^\prime_1\,(\partial K/\partial\pi^L_2)
-\pi^\prime_2\,(\partial K/\partial\pi^L_1)-
\langle\dif V|\bfe_3^L\rangle$. The general solution $|\bX_G\rangle$ of {\bf(\ref{vijf11})}, \underline{on $\cM_2$} , still contains two arbitrary functions $\zeta_1$ and $\zeta_2$ :
\be\label{vijf15}
(X_G)=
\zeta_1\,\left(\begin{array}{c}
1 \\ 0 \\ 0 \\ 0 \\ 1/\tau \\ -\pi^\prime_2 \end{array}\right) 
+\zeta_2\,\left(\begin{array}{c}
0 \\ 1 \\ 0 \\ -1/\tau \\ 0 \\ +\pi^\prime_1 \end{array}\right)
+\left(\begin{array}{c}
\partial K/\partial\pi^L_1 \\ \partial K/\partial\pi^L_2 \\ \partial K/\partial\pi^L_3 \\ 
0 \\ 
0 \\ C_{23}
\end{array}\right)\ee
This vector must be tangent to $\cM_1$ and $\cM_2$. This leads to three equations
\be\label{vijf16}
\langle\dif C_1\,|\,\bX_G\rangle=0\,;\,
\langle\dif C_{21}\,|\,\bX_G\rangle=0\,;\,
\langle\dif C_{22}\,|\,\bX_G\rangle=0\ee
If these three equations determine or not the two arbitrary functions 
$\zeta_1$ and $\zeta_2\,$, will depend on the kinetic energy $K(\pi^L)$ and on the particular form of the potential $V(g)$.
If they do so, the system will have a solution. If not, they will define a tertiary constraint manifold $\cM_3$ and the analysis must proceed.
\section{Conclusions}\label{six}
In this work, we analysed the consistency of a modification of the symplectic two-form on the cotangent bundle of a group manifold. This was done in order to obtain classical, i.e. Poisson, noncommuting configuration (group) coordinates. This was achieved in the non degenerate case, with the closed two-form $\omega_L$ which is then symplectic. 
We do not address here the general quantization problem of such a system and refer e.g. to 
\cite{Ali} for a general review on quantization methods. 
It should be stressed that, whatever the quantisation scheme, any such obtained framework 
has little to do with {\it non commutative geometry}, either in the sense of A.Connes or as a quantum field theory on non-commutative spaces.
\appendix
\section{The symplectic one-cocycle}\label{sept}
\setcounter{equation}{0}
A one-cochain $\theta$ on $\LG$ with values in $\LG^\star$, on which $\LG$ acts with the  coadjoint representation {\bf k}, $\theta\in C^1(\LG,\LG^\star,{\bf k})$, is a 
linear map $\theta:\LG\rightarrow\LG^\star:\bu\rightarrow\theta(\bu)$.  
Its components are    
$\theta_{\alpha,\mu}\doteq\langle \theta(\bfe_\mu)|\bfe_\alpha\rangle$.
It is a one-cocycle, $\theta\in Z^1(\LG,\LG^\star,{\bf k})$, if its coboundary,
$(\delta_1\theta)(\bu,\bv)\doteq{\bf k}(\bu)\theta(\bv)-{\bf k}(\bv)\theta(\bu)-
\theta ([\bu,\bv])$, vanishes.
\bean
\langle(\delta_1\theta)(\bu,\bv)|\bw\rangle&\doteq&
-\,\langle\theta(\bv)|[\bu,\bw]\rangle
+\,\langle\theta(\bu)|[\bv,\bw]\rangle
-\,\langle\theta([\bu,\bv])|\bw\rangle=0\\
\langle(\delta_1\theta)(\bfe_\mu,\bfe_\nu)|\bfe_\alpha\rangle
&\doteq&
-\,\theta_{\kappa,\nu}\;{{\bf f}^\kappa}_{\mu\alpha}
+\,\theta_{\kappa,\mu}\;{{\bf f}^\kappa}_{\nu\alpha}
-\,\theta_{\kappa,\alpha}\;{{\bf f}^\kappa}_{\mu\nu}=0
\eean
The one-cocycle $\sigma$ is called symplectic if $\Sigma(\bu,\bv)\doteq\langle\sigma(\bu)|\bv\rangle$ 
is antisymmetric, $\Sigma(\bu,\bv)=\,-\,\Sigma(\bv,\bu)$ or 
$\Sigma_{[\alpha\mu]}\doteq\sigma_{\alpha,\mu}=-\,\sigma_{\mu,\alpha}\,$. 
Any antisymmetric $\Theta$ defined in terms of $\theta\in C^1(\LG,\LG^\star,{\bf k})$ as 
$\Theta_{[\alpha\beta]}=\theta_{\alpha,\beta}$ is actually a 2-cochain on $\LG$ with values in {\bf R} and trivial representation : $\Theta\in C^2(\LG,{\bf R},{\bf 0})$. Furthermore, when $\theta \in Z^1(\LG,\LG^\star,{\bf k})$, $\Theta$ is a 2-cocycle of $Z^2(\LG,{\bf R},{\bf 0})$  :
\[
(\delta_2\Theta)(\bu,\bv,\bw)\doteq\,-\,\Theta([\bu,\bv],\bw)\,+\,\Theta([\bu,\bw],\bv)\,-\,\Theta([\bv,\bw],\bu)=0\]
\be\label{zeven1}
(\delta_2\Theta)(\bfe_\alpha,\bfe_\beta,\bfe_\gamma)\doteq
\,-\,\Theta_{\kappa\gamma}\,{{\bf f}^\kappa}_{\alpha\beta}
\,+\,\Theta_{\kappa\beta}\,{{\bf f}^\kappa}_{\alpha\gamma}
\,-\,\Theta_{\kappa\alpha}\,{{\bf f}^\kappa}_{\beta\gamma}=0\ee
In general let 
$\Theta=\frac{1}{2}\,\Theta_{\alpha\beta}\,\ep^\alpha\wedge\ep^\beta\in\Lambda^2(\LG^\star)$, obey the cocycle condition {\bf(\ref{zeven1})}. Acting with 
${L^\star}_{g^{-1}|g}$ yields the left-invariant two form :
\be\label{zeven2}
\Theta_L(g)\doteq{L^\star}_{g^{-1}|g}\;\Theta=
\frac{1}{2}\,\Theta_{\alpha\beta}\;\ep^\alpha_L(g)\wedge\ep^\beta_L(g)\ee
Using the cocycle relation and the Maurer-Cartan structure equations,   
it is seen that $\Theta_L(g)$ is a closed left-invariant two-form on $G$.\\ 
When $\LG$ is semisimple, $\Theta$ is exact. Indeed, the Whitehead lemmas state that $H^1(\LG,\bR,{\bf 0})=0$ and $H^2(\LG,\bR,{\bf 0})=0$. In particular, $\Theta\in B^2(\LG,\bR,{\bf 0})$ 
is a coboundary and there exists an element $\xi$ of $C^1(\LG,\bR,{\bf 0})\equiv\LG^\star$ such that 
$\Theta(\bu,\bv)=(\delta_1(\xi))(\bu,\bv)=-\,\xi([\bu,\bv])$ or 
\be\label{zeven3}
\Theta_{\alpha\beta}=-\,\xi_\mu\;{{\bf f}^\mu}_{\alpha\beta}\ee
The constant vector $\xi\in T^\star(\LG)$ is the analogue of a magnetic field in the abelian case $G\equiv \bR^3$.


\begin{thebibliography}{99}
\bibitem{vanhecke}
F.J.Vanhecke, C.Sigaud and A.R.da Silva, 
{\it Noncommutative Configuration Space. Classical and Quantum Mechanical Aspects},
arXiv:math-phys/0502003 and
Braz.J.Phys.{\bf 36},no IB,194(2006)
\bibitem{Connes}
A.H.Chamseddine and A.Connes, {\it The Spectral Action Principle},\\ 
Commun.Mat.Phys.{\bf186},731(1997)
\bibitem{Douglas}
M.R.Douglas and N.A.Nekrasov, {\it Noncommutative Field Theory},\\ 
 Rev.Mod.Phys.{\bf73}, 977(2001)
\bibitem{Duval}
C.Duval and P.A. Horv\'{a}thy, {\it The exotic Galilei group and the "Peierls substitution"}, 
Phys.Lett.{\bf B 479},284(2000)
\bibitem{Horvathy}
P.A. Horv\'{a}thy, {\it The Non-commutative Landau Problem},\\ 
Ann.Phys.{\bf 299},128(2002)
\bibitem{Plyushchay}
P.A. Horv\'{a}thy and M.S.Plyushchay, 
{\it Anyon wave equations and the noncommutative plane}, Phys.Lett.{\bf B 595},547(2004)
\bibitem{Abraham}
R. Abraham and J.E. Marsden, {\it Foundations of Mechanics}, \\ Benjamin,1978
\bibitem{Souriau}
J-M. Souriau, {\it Structure des syst\`{e}mes dynamiques},Dunod,1970.
\bibitem{Paulette}
P.Liberman and Ch-M.Marle, 
{\it Symplectic Geometry and Analytical Mechanics}, D.Reidel Pub.Comp.,1987
\bibitem{Azcar}
J.A. de Azc\'{a}rraga and J.M.Izquierdo, 
{\it Lie groups, Lie algebras, cohomology and some applications in physics}, 
Cambridge Univ.Press,1998. 
\bibitem{GNH}
M.J. Gotay, J.M. Nester and G. Hinds, {\it Presymplectic manifolds and the Dirac-Bergmann theory of constraints}, J.Math.Phys.{\bf 19},2388(1978).
\bibitem{Ali}
S.Twareque Ali and Miroslav Engli\v{s}, 
{\it Quantization Methods: A Guide for Physicists and Analysts}, 
Rev.Math.Phys.{\bf 17},381(2005).
\end{thebibliography}
\end{document}